# Experimental study of the structure of chalcogenide glassy semiconductors in three-component systems of Ge-As-Se and As-Sb-Se by means of NQR and EPR spectroscopy

O. N. Bolebrukh[1], N. Ya. Sinyavsky[2], I. P. Korneva[2], B. Dobosz[3], M. Ostafin[3], B. Nogaj[3], R. Krzyminiewski[3]

[1] Laboratory of Magnetic Resonance, Institute of Physics and Technology, Immanuel Kant Baltic Federal University, A. Nevski 14, 236041, Kaliningrad, Russia
[2] Division of Physics, Department of Radio Engineering, Baltic State Academy of Fishing Fleet, Molodiozhnaya 6, 236029, Kaliningrad, Russia
[3] Division of Medical Physics, Department of Physics, Adam Mickiewicz University in Poznan, Umultowska 85, 61-614 Poznan, Poland

Correspondence to be addressed to:
Dr. Irina Korneva
Department of Physics
Baltic Federal Fleet Academy
ul. Molodezhnaya, 6, Russia
tel.: +84012925117;
E-mail : ikorneva05@rambler.ru

**Abstract:**

The structure of chalcogenide glassy semiconductors in three-component systems of Ge-As-Se and As-Sb-Se has been studied by means of both NQR and EPR spectroscopy. Investigation of mutual local coordination of atoms has been performed. The results are presented for various ways of combining structural units to provide medium and long-range chemical order of chalcogenide glasses. It appears that for the Ge-As-Se system the structural transition from a two-dimensional to three-dimensional structure occurs at $\bar{r} = 2.45$.

## 1. Introduction

In recent years, non-crystalline solids have become of interest for physicists working in both fundamental research as well as applied research fields. On the one hand such systems are characterized by the lack of long-range order, i.e. strict periodicity of the arrangement of atoms in a microscopic volume. On the other hand they are characterized by the presence of short-range order, i.e., ordered distribution of the nearest neighbors for each atom. The structure, chemical and thermodynamic properties of amorphous solids are more difficult to describe in comparison to crystalline and liquid states. As a result, there are still many open questions about the properties of non-crystalline solids.

For this study the chalcogenide semiconductors of ternary Ge-As-Se and As-Sb-Se systems, the composition of which can be changed in the synthesis process, have been chosen from the large number of disordered chalcogenide materials of poorly understood inorganic origin. A characteristic feature of chalcogenide glasses is their ability to deviate significantly from the stoichiometric composition, i.e. from the respective chemical formula. For elements of a given system, it is usually impossible to obtain glass of arbitrary composition, i.e. during vitrification there are certain limits within which the amorphous materials may be prepared by rapid cooling of the melt. At the same time, samples of compositions located outside the glass forming region can be obtained, similar to production of amorphous silicon by deposition from uncondensed state. Glassy chalcogenide compounds are more thermodynamically stable as compared to amorphous silicon and germanium. Therefore, annealing usually does not lead to substantial changes in physical properties.

Nuclear quadrupole resonance (NQR) is one of the most sensitive methods to study the local structure of the material, the electron density distribution near the nucleus, the nature of the defects, the mobility of individual molecules and groups of atoms in the molecule, and etc.

NQR is used to study physical properties of solids (molecular crystallites, polymers, metals, glass, vitreous semiconductors). Applications of NQR in the study of crystals, particularly semiconductors, are based on the relationship between the structure of crystals and crystal-field gradients. In contrast to nuclear magnetic resonance, the NQR resonance frequencies are directly determined by the crystalline structure. Application of the new research methods in NQR spectroscopy (creation of the theory of the echo signal for a very broad NQR lines [1] and the geometric phase of the signal in the NQR experiments [2]) to determine the asymmetry of the electric field gradient tensor (EFG) in disordered structures allowed us to obtain more information on the structure of compounds of this type. The NQR studies have been accompanied by the electron paramagnetic resonance (EPR) study which provides information about the short-range order of the atomic arrangement.

Until now nuclear quadrupole resonance spectroscopy has not been widely used to study the structure of disordered solids [3-7]. Another radiospectroscopic method, the method of electron paramagnetic resonance spectroscopy, is used for this purpose more often [8]. These two methods provide information about the local structure, and the atomic order in amorphous solids.

The purpose of this paper is to study the structure of chalcogenide semiconductors of ternary Ge-As-Se and As-Sb-Se systems by the methods of NQR and EPR spectroscopy, to obtain information about glassy state of matter and its dependence on the composition of the CGS (chalcogenide glassy semiconductors), information about the local mutual coordination atoms, and about the ways of combining structural units providing medium and long-range chemical order.

2. **Experiment**

The process of obtaining samples of CGS for the study (Table 1) is described in [9].

**Table 1.** Investigated samples of chalcogenide glassy semiconductors: chemical formula (the ratio of concentrations), chemical formula (atomic %) and average coordination number $\bar{r}$ of the samples

| № | Chemical formula (the ratio of concentrations) | Chemical formula (atomic %) | $\bar{r}$ |
|---|---|---|---|
| **Group №1** | | | |
| 1. | $(As_2Se_3)_{0.9}(GeSe_2)_{0.1}$ | $Ge_{0.021}As_{0.375}Se_{0.604}$ | 2.417 |
| 2. | $(As_2Se_3)_{0.8}(GeSe_2)_{0.2}$ | $Ge_{0.043}As_{0.348}Se_{0.609}$ | 2.434 |
| 3. | $(As_2Se_3)_{0.7}(GeSe_2)_{0.3}$ | $Ge_{0.0608}As_{0.318}Se_{0.614}$ | 2.454 |
| **Group №2** | | | |
| 4. | $(As_2Se_3)_{0.78}(Sb_2Se_3)_{0.22}$ | $As_{0.311}Sb_{0.089}Se_{0.6}$ | 2.4 |
| 5. | $(As_2Se_3)_{0.5}(Sb_2Se_3)_{0.5}$ | $As_{0.2}Sb_{0.2}Se_{0.6}$ | 2.4 |
| 6. | $(As_2Se_3)_{0.75}(Sb_2Se_3)_{0.25}$ | $As_{0.3}Sb_{0.1}Se_{0.6}$ | 2.4 |

To describe the structure of amorphous solids, the average coordination number which is a measure of the short-range order, is typically used. The average coordination number of Ge-As-Se ($Ge_xAs_ySe_{1-x-y}$) and As-Sb-Se ($As_xSb_ySe_{1-x-y}$) glasses has been calculated by the equation from [10], where the coordination numbers for Ge, As, Sb and Se are 4, 3, 3 and 2 respectively.

NQR experiments have been carried out using a pulsed FT-NQR spectrometer type NQS-300 from MBC ELECTRONICS Company. A two-pulse Hann spin-echo pulse sequence $\frac{\pi}{2} - \tau - \pi$ was used for obtaining a NQR line. For Ge-As-Se samples the duration of the first pulse $\left(\frac{\pi}{2}\right)$ was 5 μs, the space τ between pulses was 60 μs, the duration of the second pulse (π) was 10 μs, the repetition period was 150 ms. The sampling period was 0.1 μs and the number of accumulations used was equal to 1000-2000. Measurements have been performed on $^{75}$As isotope at 77 K To record very wide NQR lines the spectrometer frequency has been varied with steps of 200 kHz. A very wide NQR spectrum is a characteristic feature of the vitreous semiconductor studied here. The NQR line width is about 20 MHz.

The reconstruction of the spectra has been carried out on the basis of integral intensities of spin-echo signals. Integrated intensity of the signal at each point was calculated using both the real and imaginary signal components in Matlab software. After that, all the spectra were fitted using a Lorentz line shape in Origin software. The expression for the sum of the two lines was used for the Ge-As-Se system, because the NQR signal for this system is observed on the nuclei of two atoms of arsenic $^{75}$As. For the As-Sb-Se system the expression for the sum of five lines was used, as for this system the NQR signal was observed, presumably, at the same time on one antimony atom nuclei $^{121}$Sb, two atoms $^{123}$Sb, and two arsenic atoms $^{75}$As.

The EPR spectra were observed with a Bruker EPR spectrometer, EMX type, working at 9 GHz frequency. The measurements were performed at 300 K and 77 K temperatures. To calculate the concentration of paramagnetic centers in investigated samples the weak pitch standard sample from Bruker was used with a concentration of $10^{13}$ spins per cm$^3$.

### 3. Analysis of the NQR and EPR spectra of Ge-As-Se and As-Sb-Se glassy semiconductors

Three-component chalcogenide glasses are much less studied than two-component glassy compositions. In order to make it possible to study the local environment of the nuclei by NQR method the compound was synthesized, in which the silicon atom without quadrupole moment was replaced by an atom of antimony in the nuclei of which ($^{121}$Sb and $^{123}$Sb) NQR could be observed.

Glasses of these systems are found in a very wide glass forming region with an average coordination number from 2 for pure selenium, up to 3.3 for compounds with a high concentration of As and Ge.

As selenium is the main piece of glass in these glasses, it is convenient to consider the local structure of a three-component system based on the structural elements of the corresponding binary systems of Ge-Se and As-Se.

According to the model of the formal valence environment, atoms with a large number of valence electrons are forming an anion sublattice.

Thus, the system's $Ge_xAs_ySe_{1-x-y}$ are presented by bonds Ge-Se and As-Se.

In a region rich with Se-atoms, the atoms of Ge and As form tetrahedral and pyramidal structural cells, respectively. In these cells branching and formation of intermolecular bonds between Se chains occurs.

When the compounds of Ge-As-Se are rich in Se atoms, associated with Ge or As atoms, which increases $\bar{r}$, it seems to be accompanied by the appearance of Ge-Ge and As-As bonds, i.e. the effects of the formation of clusters are possible.

$^{75}$As NQR spectra and their frequencies and linewidths for chalcogenide semiconductors of Ge-As-Se different compositions are presented in our previous work [11]. The NQR and EPR spectra of $^{75}$As, $^{121}$Sb and $^{123}$Sb and their frequencies and the width of the NQR lines, the EPR parameters for the chalcogenide semiconductors of different compositions of As-Sb-Se are presented in our paper [12].

The results of our study of the NQR spectra for the Ge-As-Se samples show, that change in their chemical composition leads to significant differences in the NQR spectra.

Study of the dependence of internal friction in the Ge-As-Se glasses on the average coordination number in work [13] has concluded that as $\bar{r}$ increases to a value of 2.27-2.56, the transition occurs from a one-dimensional to two-dimensional structure such as $As_2Se_3$.

The main structural units in the investigated glasses are pyramidal ($As(Se_{1/2})_3$) blocks. In the spectra of glasses a, b and c (Fig. 2) the two peaks are clearly seen, which may indicate that the structural units $As_2Se_3$ form elements that are characteristic of the crystalline structure.

According to the NQR data [14,15], in the crystal of $As_2Se_3$ $^{75}$As arsenic atom has two non-equivalent positions that differ greatly in the asymmetry parameter of the EFG tensor. In this crystal there are three selenium atoms near one of the arsenic atoms, and there are five selenium atoms near another arsenic atom, i.e. they have different coordination numbers. This leads to two different quadrupole coupling constants e2Qq/h. The asymmetry parameter is bigger for atoms with a large coordination number, which indicates the low symmetry of its environment. Quadrupole coupling constant e2Qq/h for this atom is less.

The large width of the NQR spectrum is caused by the spread of the electric field gradient (EFG) caused by the spread of bond angles in the unit cells of $As_2Se_3$ and $Sb_2Se_3$ in the glassy state. The NQR frequency of Ge-As-Se systems decreases with increasing concentrations of arsenic. This effect was not observed in the As-Sb-Se system. The position of antimony, arsenic and selenium in the periodic table of elements defines the covalent nature of the interaction between the atoms [16]. Based on previous studies it can be assumed that the present systems consist of complex structures. The main structural units in these glasses are $As_2Se_3$ or $Sb_2Se_3$, the description of which can be found, for example, in [17,18]. The results of our study suggest that the electric field gradients on the nuclei of As and Sb in the As-Se-Sb systems and crystalline $As_2Se_3$ and $Sb_2Se_3$ are very close.

4. **The interpretation of the experimental results of the CGS study and their discussion**

The interpretation of the experimental results of the CGS research is carried out on the basis of assumptions about clustering structure, as discussed above. Publications [11,12] include the NQR spectra and spectral parameters of $^{75}$As, $^{121}$Sb and $^{123}$Sb (T = 77 K) glassy systems of Ge-As-Se and As-Sb-Se in the frequency range of 48-70 MHz.

For the As-Sb-Se system, the NQR spectra indicate that besides the lines from the two nonequivalent positions of arsenic $^{75}$As nuclei, there are also the lines observed from the nuclei $^{121}$Sb and $^{123}$Sb, which complicates NQR spectrum itself and its interpretation.

Based on the NQR spectra $^{75}$As, $^{121}$Sb and $^{123}$Sb of crystalline $As_2Se_3$ and $Sb_2Se_3$, assuming that the lines are shifted but not entangled, a line of $v_1$ can be attributed to the 3/2-5/2 (or 5/2-7/2) transition of nucleus $^{123}$Sb, line $v_2$ and $v_4$ - to two non-equivalent positions of $^{75}$As nucleus, and the line $v_3$ to the 3/2-5/2 transition of $^{121}$Sb nucleus.

The graph illustrating the dependence of the NQR frequency shift on the concentration of proposed structural units $Sb_2Se_3$ is presented in Fig. 1. The value of the electric field gradient on the resonating nuclei increases with increasing concentration of isomorphic As$_2$Se$_3$, thereby the resonant frequency increases as well. Due to the ambiguous correlation between the lines in the spectra of As-Sb-Se glass, it becomes impossible to determine the asymmetry parameters by our method, because this may lead to incorrect results and conclusions. For an unambiguous correlation between the lines in these samples more research is needed with the use of dual-frequency excitation of the antimony nuclei, or with the use of selective excitation of the arsenic nuclei by radio frequency field with circular polarization.

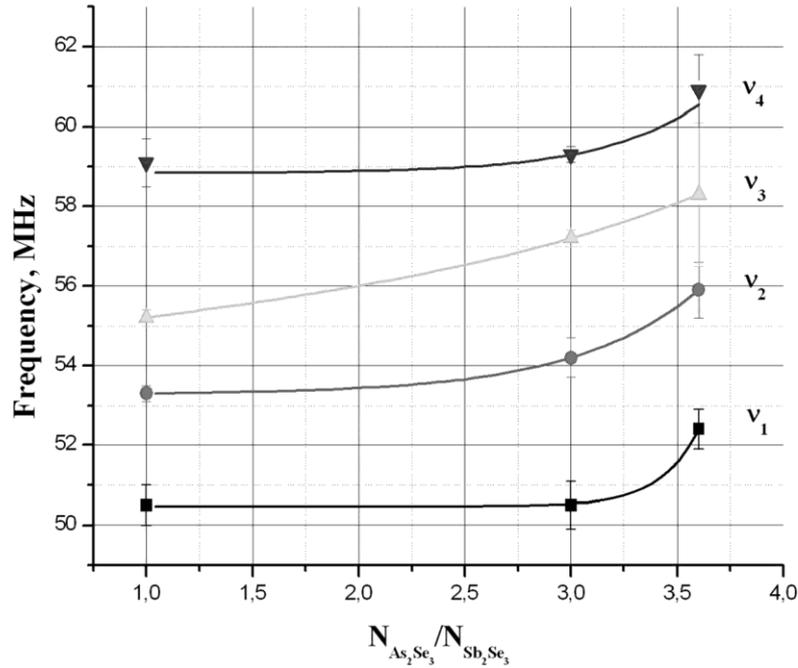

Fig. 1. The dependence of NQR frequency shift for As-Sb-Se system on the concentration of structural units of Sb$_2$Se$_3$.

As shown in [9], the asymmetry parameter of the EFG tensor is sensitive to changes in the average coordination number, i.e., to the immediate environment of arsenic. It has been found that function $\eta(\bar{r})$, exhibits a maximum at r = 2.425 which probably indicates a change in the glass structure.

Table 2 shows the frequency and width of NQR $^{75}$As in Ge-As-Se glasses (T = 77 K) from the works [6, 11].

Table 2. Frequencies and widths of the NQR lines $^{75}$As in Ge-As-Se glasses (T = 77K).

| Glass compositions | $\bar{r}$ | NQR freaquency (linewidth) | |
|---|---|---|---|
| | | $\nu_1$, MHz ($\Delta\nu_1$, MHz) | $\nu_2$, MHz ($\Delta\nu_2$, MHz) |
| Ge$_{0.021}$As$_{0.375}$Se$_{0.604}$ | 2.417 | 51,5±0,2 (2,1±0,7) | 58,4±0,3 (9,3±1,1) |
| Ge$_{0.043}$As$_{0.348}$Se$_{0.609}$ | 2.434 | 54,4±0,1 (2,9±0,5) | 59,2±0,1 (6,4±0,4) |
| Ge$_{0.0608}$As$_{0.318}$Se$_{0.614}$ | 2.454 | 55,1±0,7 (6,9±2,1) | 60,9±0,4 (5,8±1,1) |
| Ge$_{0.2222}$As$_{0.2222}$Se$_{0.5555}$ | 2.670 | 61,0±0,7 (7,8±2,0)* | 71,0±0,7 (14±2,0)* |
| Ge$_{0.33}$As$_{0.12}$Se$_{0.55}$ | 2.780 | 66,0±0,7 (9,5±2,0)* | 73,5,0±0,7(7,5±2,0)* |

*it is taken from [6].

Fig. 2 shows the dependence of the $^{75}$As NQR frequency on the average coordination number $\bar{r}$ in a disordered Ge-As-Se system. As it can be seen from the figure, there is a good correlation between the NQR frequency and coordination number (linear dependence). With the increasing $\bar{r}$ value of the EFG on the nuclei increases, this, in its turn, leads to an increase in the NQR frequencies.

In addition CVS are characterized by an average coordination number, which is the radius of the first coordination sphere and is equal to the shortest distance between neighboring atoms. Fig. 3 shows the change in the width of the NQR lines, as function of the average coordination number of glassy Ge-As-Se. It should be noted that as the line width $v_1$ increases with average coordination number, the line width $v_2$, on the contrary, decreases.

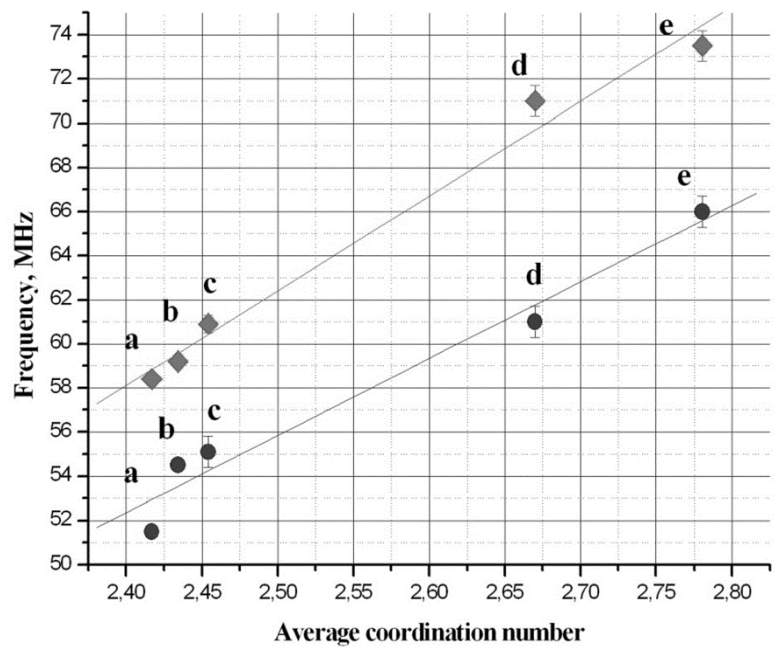

Fig. 2. $^{75}$As NQR frequency dependence on the average coordination number in a disordered Ge-As-Se system.

Fig. 4 shows the dependence of the integrated intensity of the $^{75}$As NQR lines on the average coordination number. The rhombs correspond to the first non-equivalent positions of arsenic in the structural unit of $As_2Se_3$, circles correspond to the second non-equivalent positions of arsenic.

Figs. 5 and 6 show the dependences of the displacement of the NQR frequencies on the concentration of $GeSe_2$ and the change in the widths of the NQR lines on the concentration of these clusters.

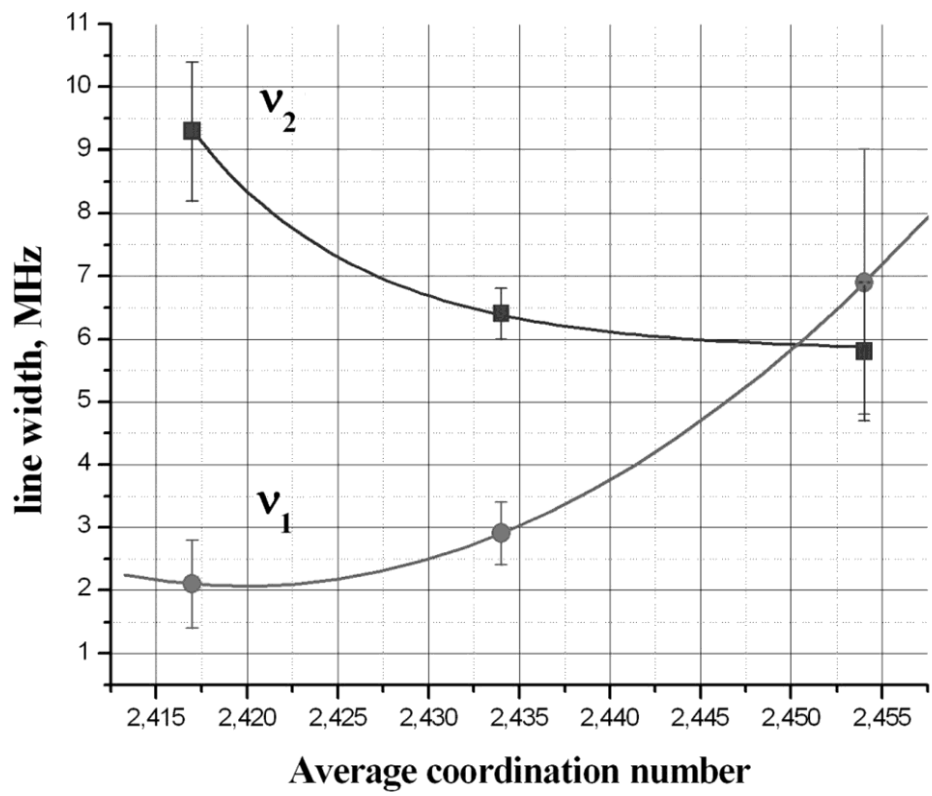

Fig. 3. Change in the width of the $^{75}$As NQR lines in Ge-As-Se glasses vs. the average coordination number.

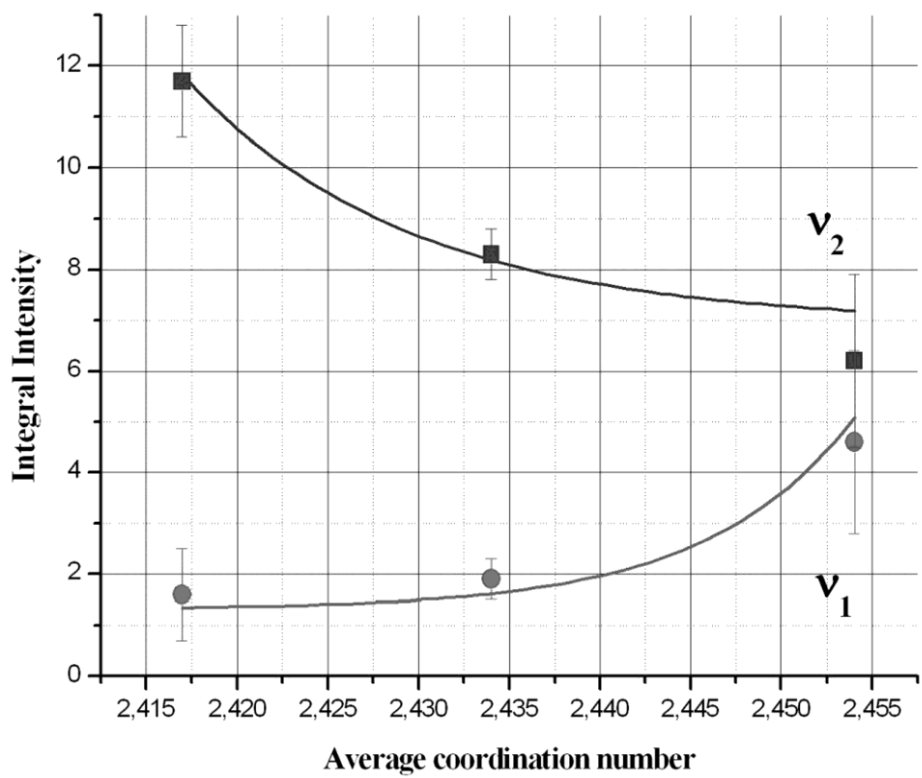

Fig. 4. The dependence of the integrated intensity of the $^{75}$As NQR lines on the average coordination number in the Ge-As-Se system.

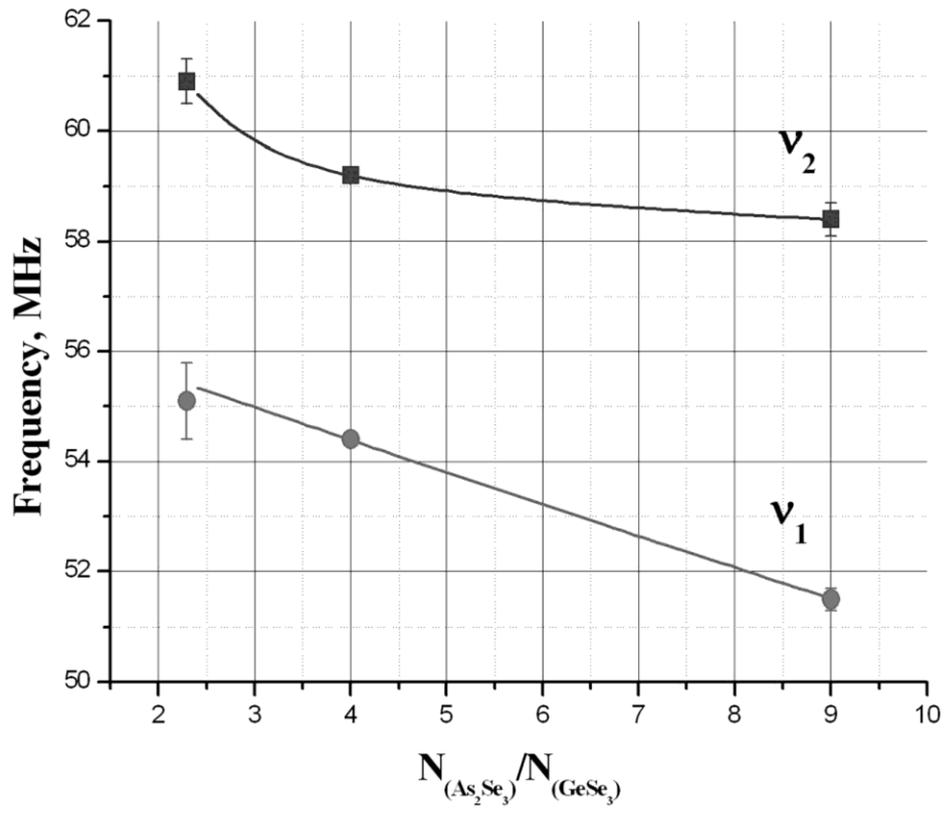

Fig. 5. $^{75}$As NQR frequency shift in the Ge-As-Se, as function of the GeSe$_2$ concentration.

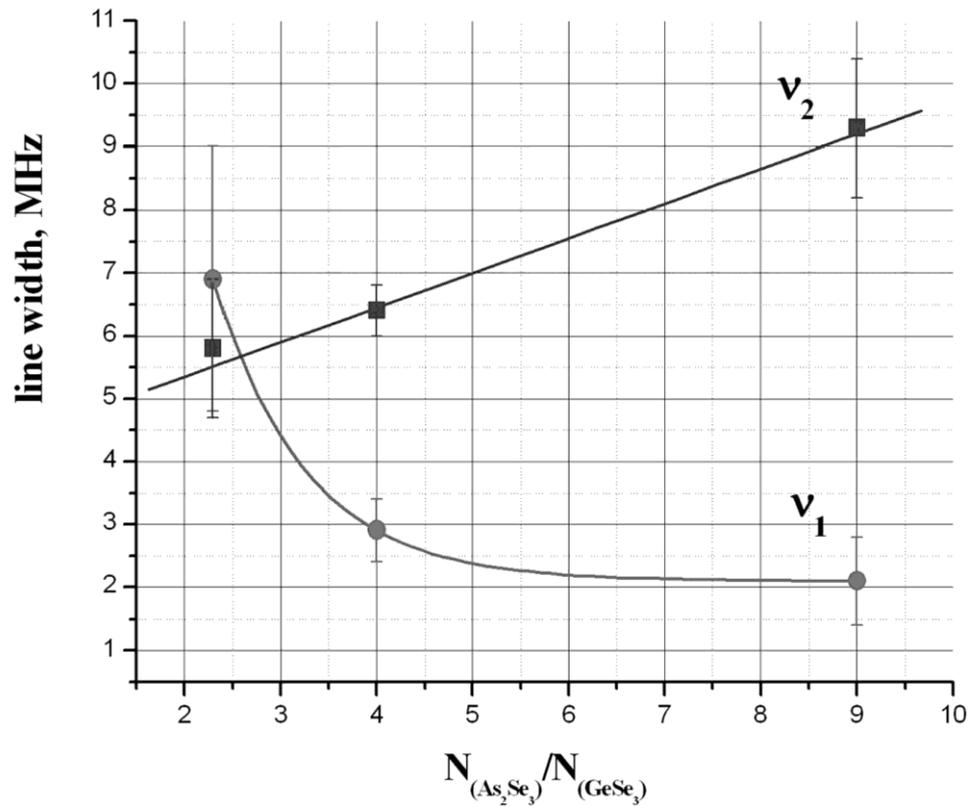

Fig. 6. Change in the width of the NQR lines in the Ge-As-Se, as function of the GeSe2 concentration. .

As it can be seen from Fig. 4, the integrated intensities of the NQR lines for the two non-equivalent positions of arsenic in the $As_2Se_3$ cluster are approximately the same at $\bar{r} = 2.45$.

As a rough analogy with grid glass crystalline lattice, let us consider the defects and their impact on the NQR spectra. As it is well known, increasing concentration of the defects in the crystalline lattice leads to observation of the decrease in intensity of the absorption line broadening NQR. Impurity molecules alter the electric field gradient on the neighboring nuclei and form local regions of tension. Each impurity molecule causes a shift of the nuclei NQR frequencies in some sphere of radius $r_C$. This is a critical region which means that all the resonating nuclei included in this region will not contribute to the center of the absorption line, which reduces the intensity of the line. At a certain concentration of impurities the function of the intensity of the NQR lines on the impurity concentration exhibits an inflection point [19], which means that the number of critical spheres has become so large that they began to overlap. This helps to slow down the decline in the intensity of the absorption line.

At the same time with the decline in the intensity and broadening of the absorption line, with introduction of impurities, the frequency shift is observed, which depends on many factors. The sign of the shift (positive or negative), in some cases, allows to determine the type of impurity. If impurity molecule is larger than the base one, then the shift is obtained in the direction of lower frequencies. Oppositely the small size impurities cause a shift towards higher frequencies.

Based on the dependence shown in Fig. 4 one can conclude that the number of the $GeSe_2$ clusters in the neighborhood of the first non-equivalent position of arsenic is decreasing with the increase of coordination number and increasing near the second unequal position of arsenic. This is because $\nu_1$-line intensity increases with the average coordination number, and the $\nu_2$ -line intensity decrease. It could also be noted that in both cases the value of the integrated intensity of the lines approaches the same value as the average coordination number approaches $\bar{r} = 2.45$. The same can be seen in the graph of change in the width of the $^{75}$As NQR lines in the Ge-As-Se glasses vs. the average coordination number (Fig. 3). It could also confirm the assumption of a structural transition at $\bar{r} = 2.45$, that is the transition of Ge-As-Se from two-dimensional to three-dimensional structure [20].

For the studied samples of the As-Sb-Se system, the EPR signal was observed as a single broad isotropic line with the absence of the fine structure. According to the research performed by other authors for similar systems, the concentration of so-called dark centers of most of the samples was about $10^{12}$ - $10^{16}$ cm$^{-3}$. The EPR spectra analysis of As-Se-Sb glasses shows that as temperature decreases from 300 to 77 K, the g-factor increases from 2.15 to 2.65, while the line width almost stays unchanged (about 954 Gs at 300 K and about 1069 Gs at 77 K). At the same temperature the intensity of the EPR spectrum of $(As_2Se_3)_{0.5}(Sb_2Se_3)_{0.5}$ is higher than the intensity of $(As_2Se_3)_{0.78}(Sb_2Se_3)_{0.22}$, which indicates a greater number of paramagnetic centers in the first sample. On the other hand, for the same compounds the line width increases inversely proportional to the content of arsenic and for $(As_2Se_3)_{0.5}(Sb_2Se_3)_{0.5}$ the line width is less than that for $(As_2Se_3)_{0.78}(Sb_2Se_3)_{0.22}$. It can be seen that with the increase in arsenic content, for the $(As_2Se_3)_{0.5}(Sb_2Se_3)_{0.5}$ sample , where the levels of arsenic are lower compared to other samples and the concentration of paramagnetic centers is higher, the line is broadened. For $(As_2Se_3)_{0.75}$ $(Sb_2Se3)_{0.25}$ and $(As_2Se_3)_{0.78}(Sb_2Se_3)_{0.22}$ samples an additional line, with g ≈ 2.05 and line width ΔB ≈ 100 Gs, is observed in the EPR spectra.

As it follows from the earlier work [21], one can assume that this line matches the paramagnetic center of selenium. The spectrum of the $(As_2Se_3)_{0.75}(Sb_2Se_3)_{0.25}$ glass exhibits a broad line of irregular shape with width of 1,016 Gs, for which g = 2.20 at T = 300 K. The increase in the line width with increasing levels of arsenic in the samples indicates a structural transformation of the glass.

Using the EPR spectra one can observe the effect of "aging" for CGS of As-Se-Sb. Figure 7 illustrates the evolution of the EPR spectra of As-Se-Sb CGS over time (7 months). The effect of "aging" of the CGS of As-Se-Sb is manifested in the broadening of the spectral lines of the EPR, increasing their intensity, and in the appearance of additional lines in the $(As_2Se_3)_{0,78}$ $(Sb_2Se_3)_{0,22}$ sample. The appearance of the narrow line is apparently caused by partial crystallization of the sample with time. This is confirmed by an experiment with the rotation of the ampoule with a sample in the cavity of the spectrometer, in which the shape of the EPR changes slightly.

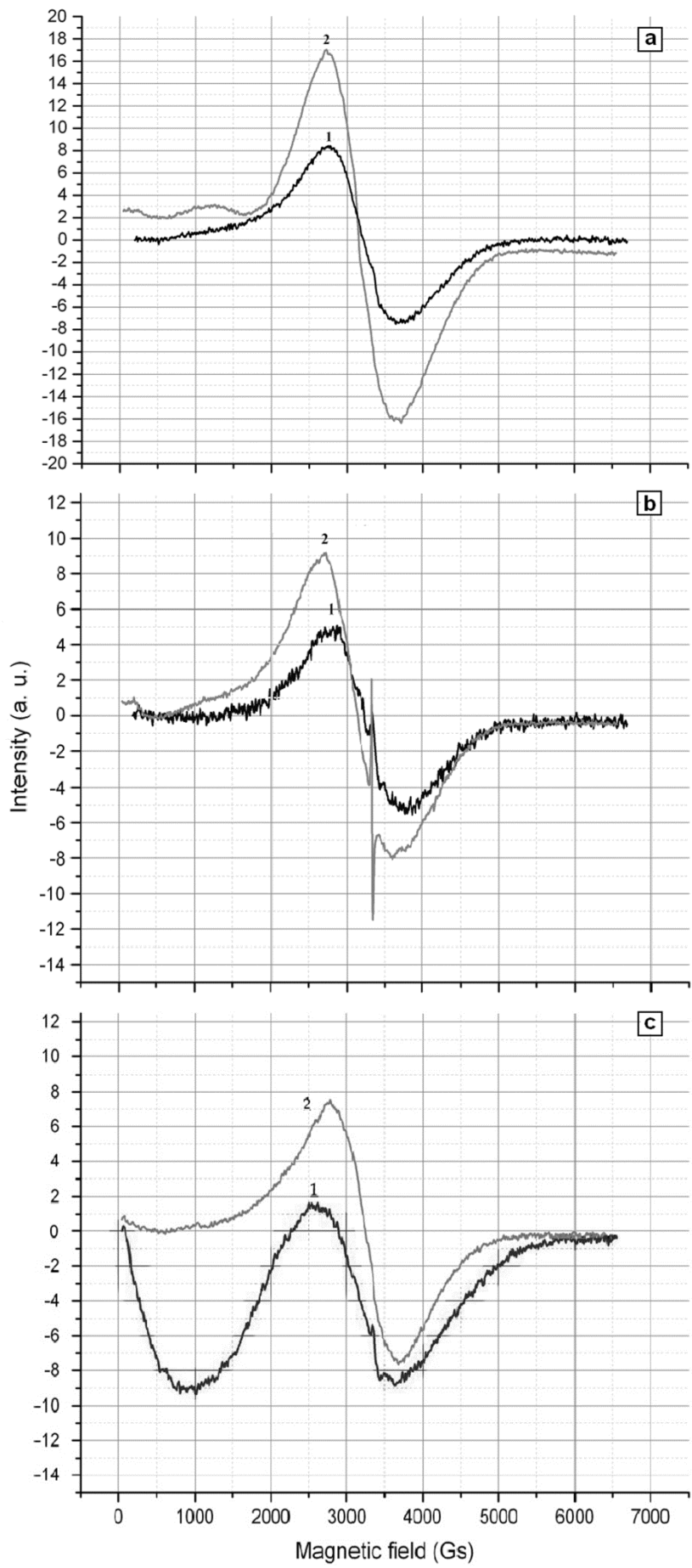

Fig. 7. The EPR spectra of CGS a - $(As_2Se_3)_{0.5}(Sb_2Se_3)_{0.5}$, b - $(As_2Se_3)_{0.78}(Sb_2Se_3)_{0.22}$ and c - $(As_2Se_3)_{0.75}(Sb_2Se_3)_{0.25}$, measured in October 2011 (1), and May 2012 (2).

## 5. Conclusion

The present study of the $^{75}$As, $^{121}$Sb and $^{123}$Sb NQR and ESR spectra of chalcogenide glassy $As_xSb_ySe_{1-x-y}$ and $Ge_xSb_ySe_{1-x-y}$ compounds, allows to establish the connection between the NQR frequencies and the line widths for $^{75}$As, $^{121}$Sb and $^{123}$Sb nuclei with a local structural order of CGS. It also suggests that clusters are forming the structure of the glasses studied.

This research produces new data for decoding of the complex multicomponent disordered structures that define short range order of atoms forming the clusters in amorphous structures. It also indicates how the structure of these substances affects their physical properties, the possibility of synthesis of structures with predictable properties, and, at the end, how to develop new technical devices based on them.